\documentclass[aps,twocolumn,prl,showpacs]{revtex4}

\begin{document}
\title{Comment on ``Probing vortex unbinding via dipole fluctuations''}
\author{S. E. Korshunov}
\affiliation{L. D. Landau Institute for Theoretical Physics,
Kosygina 2, Moscow 119334, Russia}
\date{December 5, 2006}

\begin{abstract}
We demonstrate that the method suggested by Fertig and Straley [Phys. Rev.
B {\bf 66}, 201402 (2002)] for the identification of different phases in
two-dimensional $XY$ models does not allow one to make any unambiguous
conclusions and make a tentative proposal of another approach to this
problem.
\end{abstract}


\maketitle

Recently, Fertig and Straley \cite{FS} have proposed a new method for the
identification of different phases in two-dimensional $XY$ models. They
have introduced the so-called extended dipole moment \cite{def}
$P^\alpha_{\rm ext}$ (where $\alpha=x,y$), which generalizes the notion of
vortex gas dipole moment for the systems with periodic boundary
conditions, and have suggested studying the probability of large
fluctuations of this quantity. According to Ref. \onlinecite{FS}, ${\cal
F}(L)$, the probability of finding $P^{x}_{\rm ext}$ (or, equivalently,
$P^y_{\rm ext}$) equal to $(n+1/2)L$ (where $L$ is the linear size of the
system and $n$ is an arbitrary integer), must have different dependences
on $L$ (when $L\rightarrow\infty$) in phases with bound and unbound
vortices.

With the present Comment, we would like to point out that the asymptotic
behavior of ${\cal F}(L)$ has to be the same [${\cal F}(L)\propto 1/L$]
independent of whether the vortices are bound in pairs or free, and
therefore the analysis of ${\cal F}(L)$ does not allow one to distinguish
different phases.

Consider a phase with unbound vortices. In the first approximation, one
can assume that in such a phase the vortex positions are completely
uncorrelated. In that case, it is evident from the definition \cite{FS,def}
of $P^\alpha_{\rm ext}$ that this quantity must have uniform distribution,
and therefore ${\cal F}(L)=1/L\;$. The interaction between vortices makes
the integer values of $P^\alpha_{\rm ext}/L$ more preferable than
half-integer, and therefore leads to some decrease of ${\cal F}(L)$.

Consider now a phase in which strong interaction between vortices
binds them into small neutral pairs well separated from each other.
When typical distance between vortex dipoles is much larger than their
size, their interaction is weak, which suggests that one can neglect the
correlations between the orientations of different dipoles.
In such an approximation, the total dipole moment of the system,
$P^\alpha=\sum_{j}{p^\alpha_j}$, is the sum of $N$
independent random variables $p^\alpha_j$ ($N$ being the number of vortex
pairs), so for $N\gg 1$ and $|{P}|\ll N$ it has Gaussian distribution
with the width $\sigma\propto N^{1/2}$.
Therefore, the probability to have $P^{\alpha}=\pm
L/2$ is given by
\begin{equation}                                 \label{2}
\frac{1}{(2\pi)^{1/2}\sigma}\exp\left[-\frac{(L/2)^2}{2\sigma^2}\right]
\propto \frac{1}{L}
\end{equation}
where we have taken into account that in large systems $N=cL^2$,  $c$
being the concentration of vortex pairs. Since
$P^\alpha=P^\alpha_{\rm ext}$ (mod $L$),
Eq. (\ref{2}) also gives the dependence of ${\cal F}(L)$ on $L$.

Now we have to return to the interaction between vortex dipoles. If
it would be dependent directly on the relative orientation of the dipoles,
it would lead to suppression (or enhancement) of the fluctuations of total
dipole moment. However, in two dimensions the dipole-dipole interaction
\begin{equation}                                 \label{di}
    E({\bf p},{\bf p'})\propto
    \frac{({\bf p}{\bf p'})r^2-2({\bf pr})({\bf p'r})}{r^4}
\end{equation}
depends not on the relative orientation of two dipoles \cite{comm2}
but on their orientations with respect to the line which connects them.
This interaction does not force the total dipole moment
of a diluted gas of dipoles to become smaller (or larger),
and therefore cannot lead to suppression (or enhancement) of
its fluctuations in comparison with the case of noninteracting
dipoles.

In Ref. \onlinecite{FS}, the fluctuations of dipole moment in the phase
with bound vortices were analyzed with the help of the duality
transformation with a subsequent replacement of a discreet solid-on-solid
(SOS) model by a continuous one (with a renormalized coupling). In terms
of the original XY model, this corresponds to replacing the cosine
interaction by a harmonic one. 
Naturally, such a replacement leads to the complete suppression of vortices,
which fixes the total dipole moment at zero [as follows also from Eq. (8)
of Ref. \onlinecite{FS}]. It is clear that the disregard of
vortices makes this approach utterly unsuitable for analyzing the
fluctuations of vortex gas dipole moment.

Since we have found that the asymptotic dependence of ${\cal F}(L)$ on $L$
is the same both for noninteracting vortices and for strongly bound
vortices, one can expect than it will be the same also in all intermediate
cases. Therefore, the numerical calculation of ${\cal F}(L)$ cannot help
one distinguish the phases with bound and unbound vortices.

The observation in the numerical simulations of Ref. \onlinecite{FS} of
the situations in which ${\cal F}(L)$ decays with the increase of $L$ much
faster than $1/L$ can be explained as
a manifestation of transitional regimes in which the main contribution to
${\cal F}(L)$ is related to the formation of large dipoles
and not to the reorientation of small dipoles.
However, our analysis suggests that with a
further increase of $L$, a crossover to the regime in which the
mechanism described in this Comment is the dominant one must occur.
Note that in the case of strong coupling,
the coefficient between ${\cal F}(L)$ and $1/L$ is exponentially small in
vortex pair concentration $c$, and therefore  the
observation of the asymptotic dependence of ${\cal F}(L)$ requires
rather large values of $L$ (with $\ln L\gg 1/c$).

Possibly, a more consistent approach to the identification of different
phases in two-dimensional $XY$ models can be based on analyzing in dynamic
simulations the behavior of ``world lines" of vortices
in the three-dimensional space-time.
These world lines are in many respects analogous to vortex lines in
a three-dimensional XY model. In particular, they have to be continuous
due to the conservation of the topological charge.
It follows from the results of Ref. \onlinecite{Lebedev}
that in terms of vortex world lines, the main difference between the phases
with bound and unbound vortices consists in the absence (or existence) of
the world lines of infinite length (crossing  a ``whole sample"
in time direction). In the phase with bound vortices
all world lines have to form closed loops, whereas in the
phase with unbound vortices the concentration of infinite world lines
(which can be associated with free vortices) has to be nonvanishing.

This suggests that in the framework of a dynamic description,
the concentration of free vortices is a well defined quantity
(in contrast to thermodynamic simulations, in which it is impossible
to introduce an algorithm for counting the number of free vortices in a
given vortex configuration). Therefore, a numerical calculation of this
concentration may turn out to be a useful instrument for distinguishing the phases
with and without free vortices.

$~$

The author is grateful to H.A. Fertig
for useful exchange of opinions and to V.V. Lebedev for enlightening
discussion.



\begin{thebibliography}{99}
\bibitem{def}$P^x_{\rm ext}$ is an integer
number which is changed by $+1$
when a positive (negative) vortex jumps to the next lattice cell
to the right (to the left), and by $-1$ when a jump occurs in the
opposite direction.
\end{thebibliography}
\end{document}